\documentclass{elsart}
\usepackage[dvips]{graphicx}
\usepackage{amssymb}

\begin{document}
\begin{frontmatter}

\title{Chain Length Dependence on Folding Transition of a Semiflexible Homo-Polymer Chain: Appearance of a Core-Shell Structure}

\author[kyoto]{Yuji Higuchi}
\author[fukui]{Takahiro Sakaue}
\author[kyoto,kyoto1]{Kenichi Yoshikawa \corauthref{cor}}
\corauth[cor]{Corresponding  author. Fax: +81 75 753 3779}
 \ead{yoshikaw@scphys.kyoto-u.ac.jp} 
\address[kyoto]{Department of Physics, Graduate School of Science, Kyoto University, Kyoto 606-8502, Japan}
\address[kyoto1]{Spatio-Temporal Order Project, ICORP, JST, Kyoto 606-8502, Japan}
\address[fukui]{Fukui Institute for Fundamental Chemistry, Kyoto University, Kyoto 606-8103, Japan}

\begin{abstract}
The folding transition of single, long semiflexible polymers was studied with special emphasis on the chain length effect using Monte Carlo simulations. While a relatively short chain (10-25 Kuhn segments) undergoes a large discrete transition between swollen coil and compact toroid conformations, a long chain (50 Kuhn segments) exhibits an intrachain segregated state between the disordered coil and ordered toroid.
\end{abstract}
\end{frontmatter}

\section{Introduction}
The conformational transition of a polymer chain from a statistical coil to a condensed state has been actively studied over the past several decades.
According to the classical scenario, a polymer coil gradually shrinks into a spherical globule with a decrease in the solvent quality or an increase in the pair-wise attraction between monomers~\cite{Groseberg_Khokhlov}.
However, once the effect of the chain stiffness is taken into account, the situation changes, even qualitatively. 
For a single semiflexible polymer, the collapse transition becomes discontinuous and shows the characteristics of a disordered disperse state and an ordered condensed state~\cite{Groseberg_Khokhlov,PRL_Yoshikawa,JCP2_Sakaue,CPL_Noguchi}. This transition has been demonstrated by the single molecule observation of giant DNA molecules~\cite{PRL_Yoshikawa,CPL_Noguchi}\\
Under usual aqueous conditions, DNA possesses a rigid helical structure, which gives rise to a large persistence length $l_p \simeq 50$ nm compared to its molecular thickness $d \simeq 2$ nm.  Therefore, giant DNAs with contour length $L \gg 1$ $\mu $m is characterized as a typical semiflexible polymer chain.
There have been many numerical and theoretical studies on the folding transition of a semiflexible chain~\cite{Groseberg_Khokhlov,JCP2_Sakaue,CPL_Noguchi,JCP2_Noguchi,JCP_Ivanov,Macromol_Ivanov,JCP_Stukan,PRE_Stukan,PRE_Schnurr,PRE_Miller,JCP_Kuznetsov,JCP_Ou,Biophys_Ubbink,Biophys_Odijk,Biopolym_Vasilvskaya,JCP_Sakaue}. These studies show a variety of collapsed morphologies, such as a toroid, a rod, and a spherical globule, along with strong finite chain length effect. It has been shown that a toroid is the most typical structure as the compact state, whereas some other morphologies are also found, such as a rod, a sphere, and a composite between a toroid and rod. Although observations by computer simulation have been shown to be very powerful, the range of chain length studied so far has been rather limited; i.e., the length is too short, which is at most on the order of 10 Kuhn segments.  On the other hand, genomic DNA molecules have a contour length on the order of mm$\sim$cm, indicating that the number of Kuhn segments is on the order of $10^4$-$10^5$.  Even for phage DNAs, the number is on the order of $10^2$. Thus, studies on long semiflexible chain would have biological significance, in addition to physical interest~\cite{Adv_Yoshikawa}. On the other hand, theoretical studies usually postulate some specific morphologies and analyze their relative stabilities by a grand state or mean-filed type  approximation. Overall, our understanding of the collapse transition of long semiflexible polymers is far from complete, and it would be desirable to describe a generic scenario that could be expected for longer chains.\\
In this Letter, we report the nontrivial scenario of the collapse transition found in long semiflexible polymers based on a systematic Monte Carlo simulation.

\section{Methods}
To investigate the collapse transition of long semiflexible polymers, we carried out off-lattice Monte Carlo simulations with a Metropolis algorithm in three-dimensional space, in which the polymer is represented as $N$ beads (diameter $\sigma$) connected by springs.  The potential energy of the system is represented by the following three terms;
\begin{equation}
U_{beads}=T\sum_{i=1}^{N-1} \frac{k_{bond}}{2\sigma ^2}(\mid {\bf r_{i+1}}-{\bf r_i}\mid -\sigma)^2 
\end{equation}
\begin{equation}
U_{bend}=T\sum_{i=2}^{N-1} \frac{k_{\theta}}{2}(1-\cos\theta _i)^2 
\end{equation}
\begin{equation}
U_{LJ}=\sum_{\mid i-j\mid >1}4\epsilon \left[\left(\frac{\sigma}{\mid {\bf r_i}-{\bf r_j}\mid}\right)^{12}-\left(\frac{\sigma}{\mid {\bf r_i}-{\bf r_j}\mid}\right)^6\right]
\end{equation}
where ${\bf r_i}$ is the coordinate of the $i$th monomer, $\theta _i$ is the angle between adjacent bond vectors and $T$ is the thermal energy with the Boltzmann constant $k_B=1$.  The length and energy are measured in units of $\sigma$ and $\epsilon$, respectively. 
We set the spring constant $k_{bond}=40$ and impose a cut-off, i.e., $U_{beads}=\infty$, if $\mid {\bf r_{i+1}}-{\bf r_i}\mid > 1.15\sigma$ or $\mid {\bf r_{i+1}}-{\bf r_i}\mid <0.85\sigma$ to avoid unrealistic bond stretching.
For the bending elasticity, we choose $k_{\theta } =60$, which corresponds to a constant persistence length $l_p\simeq 10\sigma$ independent of $T$. The excluded-volume and attractive effects are included in the Lennard-Jones potential $U_{LJ}$.  We set $N=200$, 500, and 1000, which correspond to a DNA chain of about 3, 7.5, and 15 kilo base pairs, respectively.\\
For the trial motion, we adopted a slithering-snake motion, in which an end monomer is removed and attached to the other end of the chain, as well as the usual random-hopping method.
The introduction of a slithering-snake motion makes sampling in the configurational space very efficient and enables us to obtain statistics even for long chains. This is particularly useful in transition regions, which allows for transitions among different (meta) stable states several times in a single long simulation on the order of $10^7 \simeq 10^8$ Monte Carlo steps.
Each calculation was started from a randomly generated swollen coil conformation, and the statistical quantities were sampled every 5000 Monte Carlo steps after equilibration.
\\
We introduce two ``order-parameters" concerning the statistical quantities here.  One is the quantity of the folded fraction of monomers, and is defined as
\begin{equation}
P=\sum_{i=1}^{N} \frac{\rho (i)}{N}=\sum_{i=1}^{N} \frac{1- \prod_{\mid i-j \mid >\lambda}(1-\rho_{i,j})}{N}
\end{equation}
where $\rho_{i,j}$ is an indicator of the pair contact:
$\rho_{i,j}=1$ if the number of monomers, which satisfies $\mid {\bf r_i}-{\bf r_j}\mid <r_{c1}$, is more than 1, and otherwise $\rho_{i,j}=0 $.
Therefore, $\rho(i)$ represents a binary local density ($\rho =1$ if the $i$-th monomer has some surrounding neighbors and $\rho=0$ otherwise).
The other is following local orientational order $\eta$,
\begin{equation}
\eta =\frac{1}{2}(3 \langle \cos ^2\theta_{ij} \rangle_{local} -1) 
\end{equation}
where the bracket $\langle \rangle_{local}$ indicates averaging over the bond pairs $\mid i-j \mid > \lambda$, the spatial distance between which is less than $r_{c2}$.
In the following discussion we set $r_{c1}=2.5\sigma$, $r_{c2}=1.5 \sigma$ and $\lambda =3$.

\section{Results}
Figure 1 (a) shows the average of the normalized gyration radius $ \langle R_g \rangle /N^{\frac{1}{2}}$ as a function of the inverse temperature. The bracket $\langle \cdot \rangle$ denotes the ensemble average.
A swollen coil collapses into condensed states with a decrease in temperature, and the transition is discontinuous, as shown by the bimodal distribution of $R_g$ around the transition point. To clarify this point, we divide a total ensemble into subensembles with different states (here the coil and condensed states) and calculate the average quantity in each subensemble. 
Figure 1 (b) shows the fluctuations of the gyration radius $( \langle R_g^2 \rangle - \langle R_g \rangle ^2)/N$. A large peak is observed in the transition region, which reflects the co-existence of the coil and condensed states.
These data indicate that with an increase in the chain length, the transition temperature becomes higher and the discontinuity increases.\\
Figure 1 (c) shows the fluctuation of enthalpy $\langle E^2 \rangle - \langle E \rangle ^2$, where $E=U_{bond}+U_{bend}+U_{LJ}$, to check the change in properties upon a folding transition beside the parameters that are related to the gyration radius. For short chains ($N=500$ and 200), the temperature dependence showed only a single peak at $\epsilon /T\simeq 0.47$ and $\epsilon /T\simeq 0.56$, respectively, as shown in Fig. 1 (c). The single peak corresponds to a large discrete transition between the coil and compact states, where the latter state is characterized by the morphology of a compact toroid, which is consistent with previous studies.  In contrast, for a long chain ($N=1000$), a pair of peaks are noted at $\epsilon /T\simeq 0.41$ and $\epsilon /T\simeq 0.44$, indicating that the folding transition proceeds via two steps.
\begin{figure}
\begin{center}
\includegraphics[width=.4\linewidth]{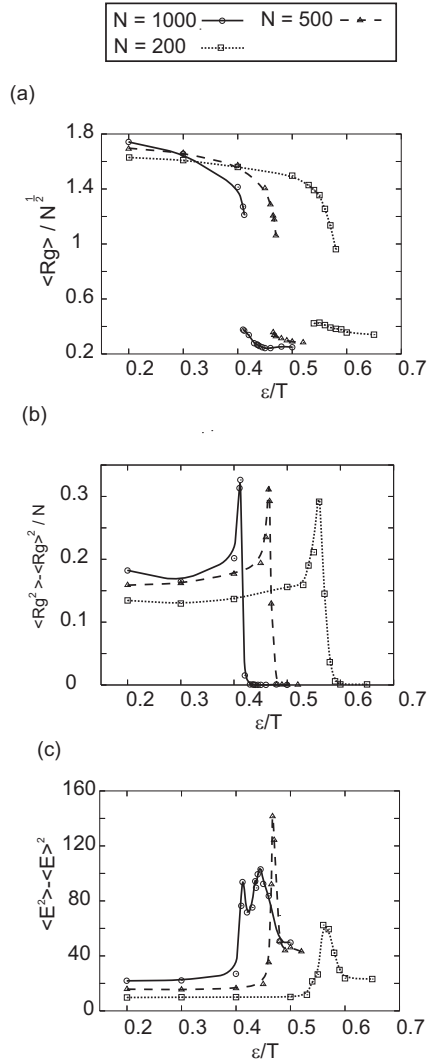}
\caption{(a): Normalized gyration radius $\langle R_g \rangle$ versus inverse temperature $\epsilon /T$. In the transition region, where the distribution becomes bimodal, each subensemble is treated separately (see the text).  The lines are simply visual guides.
(b): Fluctuations of gyration radius $(\langle R_g^2 \rangle-\langle R_g \rangle^2)/N$ versus inverse temperature $\epsilon /T$. 
(c): Fluctuations of enthalpy $\langle E^2 \rangle-\langle E \rangle^2$ versus inverse temperature $\epsilon /T$.}
\label{fig1}
\end{center}
\end{figure}\\
Three-dimensional snapshots are shown in Fig. 2 (a) at $\epsilon /T= 0.46$ (after the second peak) and in Fig. 2 (b) at $\epsilon /T= 0.43$ (between the first and second peaks).  While Fig. 2 (a) is a regular toroidal structure, the long semiflexible chain also exhibits a qualitatively different condensed state: a core-shell structure (Fig. 2 (b)), which consists of a condensed core and a coiled shell around it. The overall shape of the core is spherical, and is topologically different from a toroid. In contrast to a toroid, there is no angle to describe the features of a core-shell.
\begin{figure}
\begin{center}
\includegraphics[width=.5\linewidth]{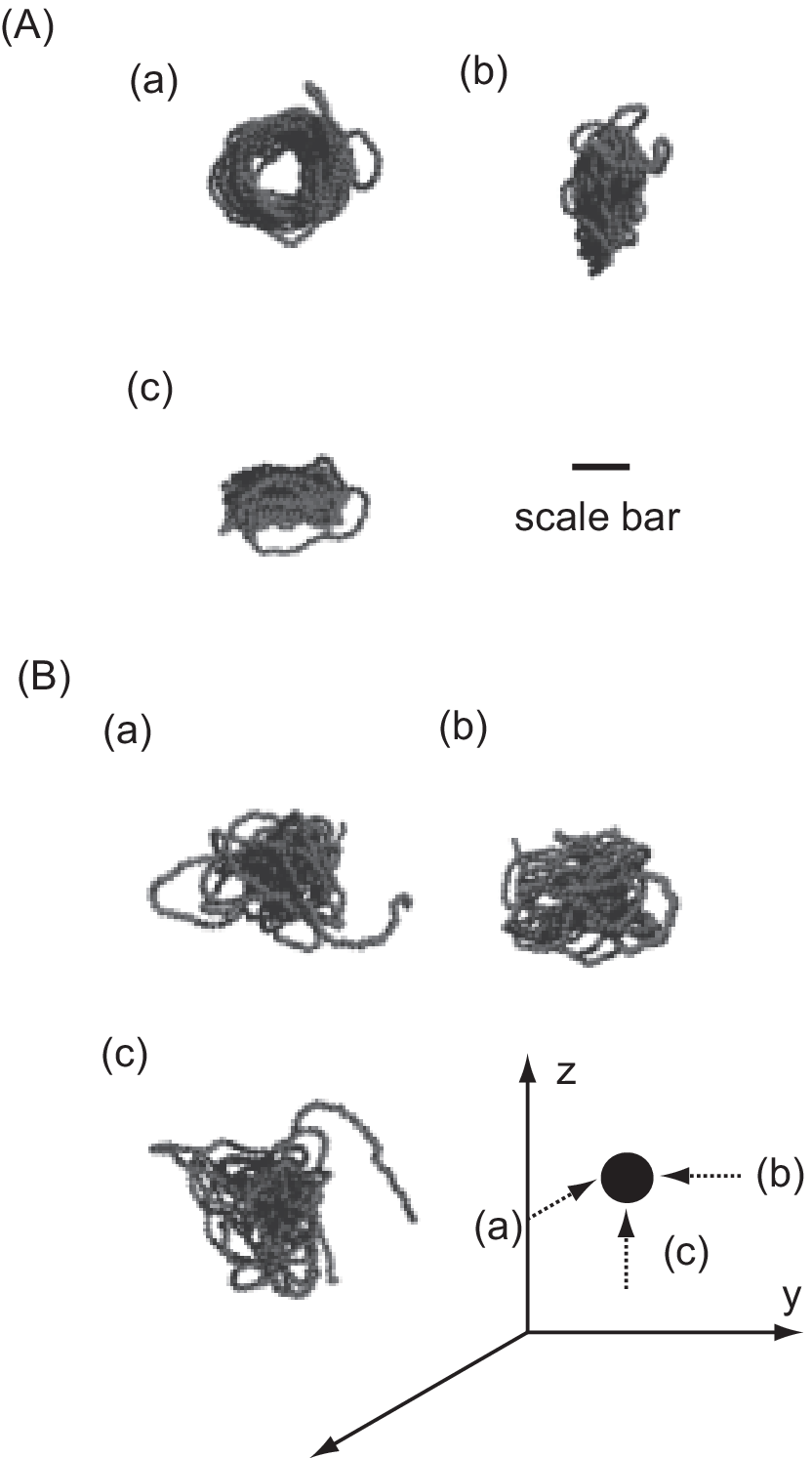}
\caption{Typical snapshots of the collapsed structure for a chain with $N=1000$. (A) a toroid at $\epsilon/T=0.46$ and (B) a core-shell at $\epsilon/T=0.43$. The bar represents the persistence length; $l_p=10 \sigma$.}
\label{fig2}
\end{center}
\end{figure}
With a decrease in temperature, the first transition of the coiled state to a core-shell is in the first peak, $\epsilon /T\simeq 0.41$, and the second transition of a core-shell to a toroid is in the second peak $\epsilon /T\simeq 0.44$.
When $\epsilon/T \simeq 0.43$, the initially prepared toroid structure spontaneously transforms into the core-shell structure, which suggests that the core-shell is a globally stable structure in this temperature range.\\
Figure 3 (A) shows the distribution of $\eta$ at three different temperatures for $N=1000$.  The degree of local orientational order clearly shows a difference between the toroid and core-shell. In the core-shell ($\epsilon /T=0.43$) and toroid ($\epsilon /T=0.46$) states, single peaks are seen near $\eta=0.2$ and $\eta=0.4$, respectively.  On the other hand, two or three peaks are observed at $\epsilon /T =0.44$, which indicates the co-existence of structures with different orientational orders.
\begin{figure}
\begin{center}
\includegraphics[width=.7\linewidth]{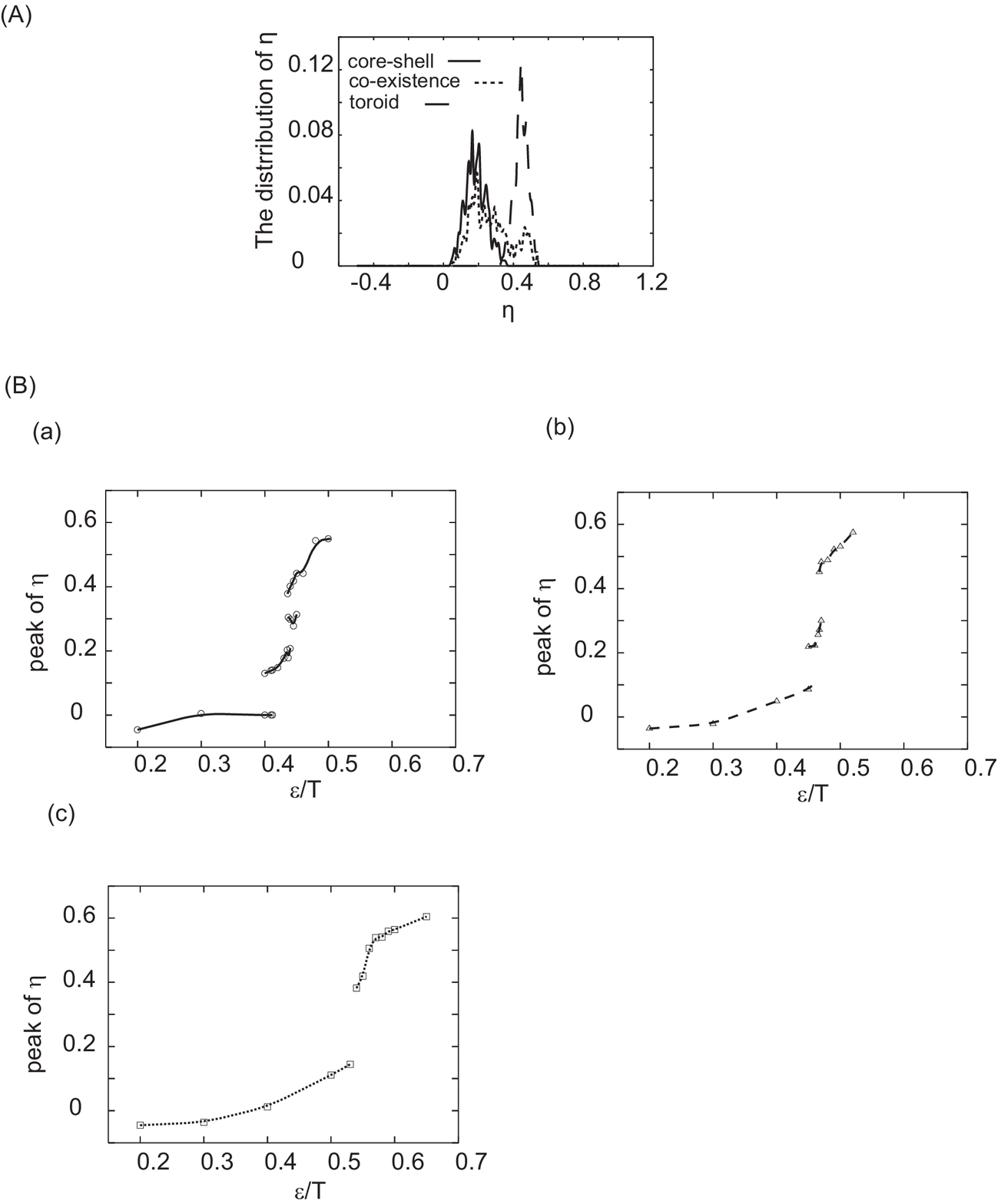}
\caption{(A)  Distribution of the local orientational order $\eta$ for a chain with $N=1000$ at various temperatures $\epsilon /T$. Solid, dashed and dotted curves correspond to $\epsilon/T=0.43$ (core-shell), $\epsilon/T=0.44$ (co-existence) and $\epsilon/T=0.46$ (toroid), respectively. (B) Peak of orientational order $\eta _{peak}$ versus inverse temperature $\epsilon /T$. In coiled state, there are no peak in $\eta$, so the averages of orientational order $\langle \eta \rangle$ are plotted.
(a): For $N=1000$, there are four states; coil $\langle \eta \rangle \simeq 0$, core-shell $\eta _{peak}\simeq 0.2$, core-shell (with toroidal core) or disk $\eta _{peak}\simeq 0.3$, and toroid $\eta _{peak}\simeq 0.4 \sim 0.6$. 
(b): For $N=500$, there are three states; coil, core-shell (with toroidal core) at $\eta _{peak}\simeq 0.2 \sim 0.3$, and toroid. 
(c): For $N=200$, there are two states; coil and toroid.}
\label{fig3}
\end{center}
\end{figure}

Figure 3 (B) shows the peaks of orientational order $\eta _{peak}$ versus inverse temperature.
For $N=1000$ (Fig. 3 (a)), the folding transition is characterized by a two-step mechanism; the first step corresponds to the collapse from a swollen coil to a core-shell, and the second involves the transition from a core-shell to an ordered toroidal structure with a decrease in temperature.  In the state when the core-shell and toroid co-exist ($\epsilon /T\simeq 0.445$), a close inspection reveals that there are three states: the core-shell, the toroid, and another with a corresponding peak in local orientational order at around $\eta _{peak}\simeq 0.3$.  Based on snapshots, the structures would be a core-shell of different type with not a spherical but a toroidal core, or a disk, which is similar to a toroid but there is no hole in the center. 
The data with $N=500$ (Fig. 3 (b)) indicate the possible occurrence of structures with $\eta _{peak}\simeq 0.3$ even at this chain length within a narrow temperature interval (perhaps as metastable states).
\begin{figure}
\begin{center}
\includegraphics[width=.4\linewidth]{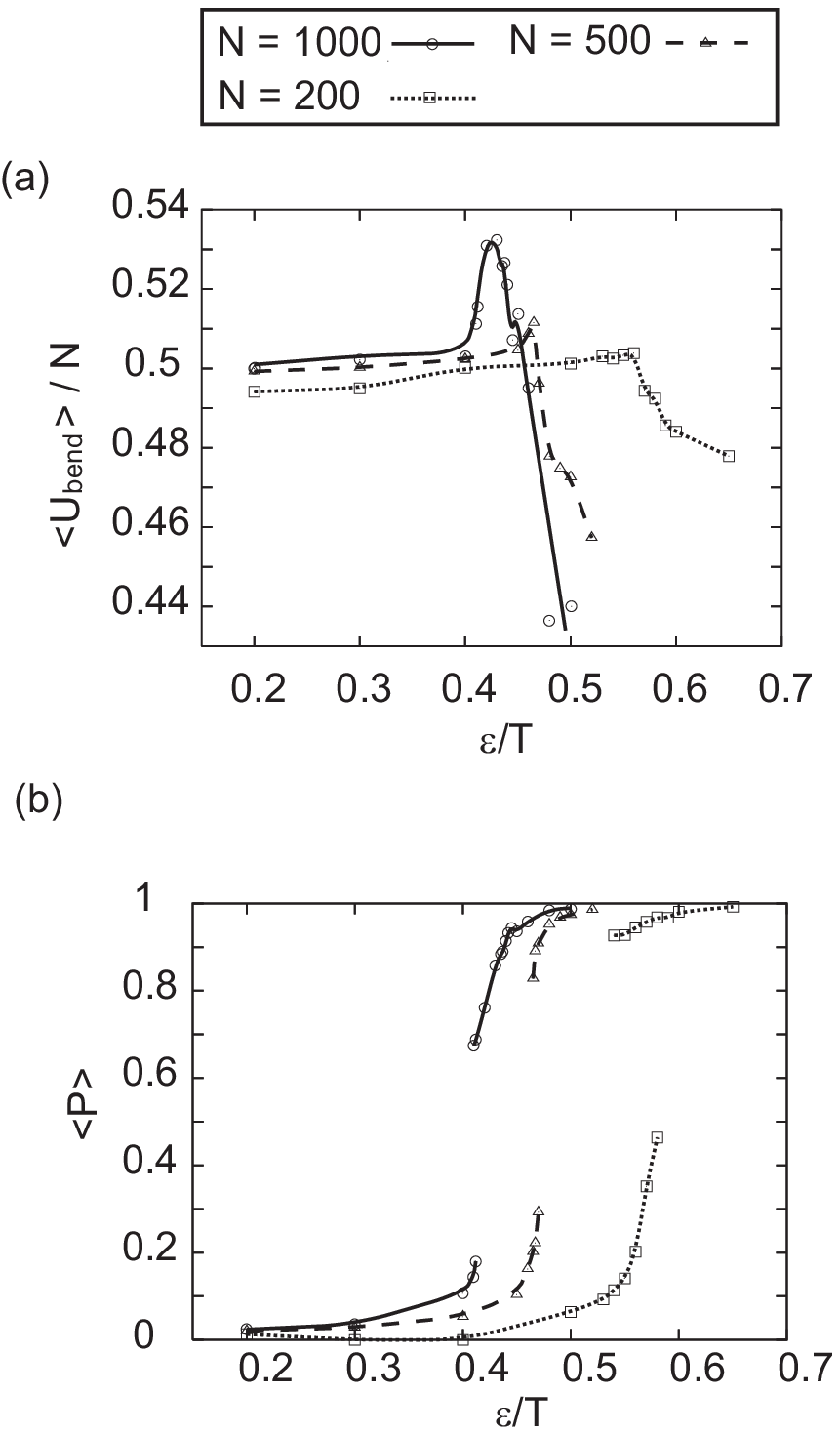}
\caption{
(a): Average bending energy $\langle U_{bend}\rangle$ versus inverse temperature $\epsilon /T$.
(b): Fraction of folded segments $\langle P \rangle$ versus inverse temperature $\epsilon /T$.}
\label{fig4}
\end{center}
\end{figure}\\
Figure 4 (a) shows the bending energy per unit $\langle U_{bend} \rangle /N$. While the toroidal structure is favorable in terms of bending, the core-shell has  a larger bending energy than a toroid and a coil.  Note also the small peak in the case of $N=500$, which again indicates the appearance of core-shell or disk structures in this region.  Figure 4 (b) shows the average fraction of folded segments $\langle P \rangle$.  $\langle P \rangle$ of the core-shell is clearly lower than that of the toroidal structure ($\langle P \rangle \simeq 1$), which indicates the partially folded state, and thus the cohesive energy of the core-shell is lower than that of the toroidal structure. These results suggest that the stability of the core-shell has an entropic origin because the core-shell has a greater enthalpy than the toroidal structure. 

\section{Discussion}
The stability of the toroidal structure has been discussed in detail by several authors~\cite{PRE_Miller,Biophys_Ubbink,Biophys_Odijk,Biopolym_Vasilvskaya,JCP_Sakaue}.
These theories claim that the toroid becomes fat with a decrease in solvent quality and/or upon lengthening of the chain, which leads to shrinkage of the toroid hole.  As a result, the toroid is assumed to transform into a disk or sphere as low temperature structures.
A rod-like structure has also been reported to be stable in some parameter regions~\cite{JCP2_Noguchi,PRE_Schnurr}.
All of these studies considered the stability of a condensed state at sufficiently low temperatures, i.e., well below the collapse transition point, where all monomers show the condensed state ($P =1$). While various morphologies have been suggested due to the finiteness of the chain length, they are implicitly assumed to be in the same ordered ``phase". \\
In contrast to the current understanding, in this study we demonstrated that a single long semiflexible polymer can assume distinct condensed ``phases" characterized by different orientational orders. Such a chain can also assume a partially collapsed state with $P <1$.
These results show that the structure of a single long semiflexible polymer can be tuned upon the folding transition.
These features may be relevant to the biological function of DNA molecules through the control of higher-order structures.\\
T4 DNA (166 kilo base pairs, about 500 Kuhn segments) and $\lambda$- DNA (48 kilo base pairs, about 150 Kuhn segments) are examples of giant DNA that are often used in experiments, and it is expected that they are long enough to exhibit a core-shell structure.
There are some experimental indications (though not conclusive) regarding the appearance of the core-shell structure as a collapsed state of these giant DNA molecules near the transition region.
First, fluorescent microscopic observations often reveal ``blurred" images of collapsed DNAs around the transition region, which is signified by the fluorescent intensity distribution as well as relatively slow Brownian motion. 
Second, distinct elastic responses in an experiment on the mechanical unfolding of $\lambda$- DNA have suggested the presence of different condensed states~\cite{PRL_Murayama}.  While the stick-release pattern observed in the deep condensed region most likely reflects the ordered toroid structure, another pattern with a long plateau in the force-distance curve which is characteristic of the shallow condensed region may be attributed to a more loose and disordered structure.
Furthermore, as more direct evidence, electron microscopic observations have shown that a single T4 DNA molecule assumes a partially folded core-shell structure and/or a loosely packed folded structure under certain conditions~\cite{Bio_Damien}.\\
The large discrete nature of the folding transition of giant DNA has been demonstrated in single DNA observation through the use of fluorescence microscopy.  Although such observations have clarified this discrete nature, the detailed microscopic structure of the product of the folding transition has only been visualized using methodologies such as electron microscopy and atomic force microscopy. Unfortunately, these methodologies only provide information on species adsorbed on a solid surface. However, some experiments indicate the presence of a loose, less ordered collapsed state compared to an ordered toroid in the shallow condensed condition, and it may be reasonable to interpret these features in terms of the core-shell structure found in the present study.

\section{Conclusion}
We studied the nature of the folding transition of a semiflexible polymer chain with changes in the contour length through Monte Carlo simulation. For relatively short chains (10 or 25 Kuhn segments), this transition is all-or-none type and produces a toroidal structure. In contrast, for long polymer chains (50 Kuhn segments), this transition is a multiple-step process, in which a swollen coil is first collapsed into a partially folded core-shell structure, and is followed by the subsequent transitions to the completely folded ordered structure upon further quenching.  Our findings, and in particular the appearance of a partially folded core-shell as a stable structure, may be unexpected based on the current understanding in this field. A statistical analysis indicated that the core-shell is an entropically stabilized state, and its appearance should be rather ubiquitous in the folding of long semiflexible chains. The fact that long semiflexible polymers possess such structural variability without a drastic change in the spatial size may have important consequences in the context of DNA function {\it in vivo}. Further theoretical and experimental studies are needed to fully elucidate the folding transition of long semiflexible polymers.
\ack
This work was supported by Japan Society for the Promotion of Science (JSPS) under a Grant-in-Aid for Creative Scientific Research (Project No. 18GS0421).


\begin{thebibliography}{00}

\bibitem{Groseberg_Khokhlov}A.Yu. Grosberg, A.R. Khokhlov, {\it Statistical Physics of Macromolecules} (American Institute of Physics, New York, 1994).
\bibitem{PRL_Yoshikawa}K. Yoshikawa, M. Takahashi, V.V. Vasilevskaya, A.R. Khokhlov, Phys. Rev. Lett. 76 (1996) 3029.
\bibitem{JCP2_Sakaue}T. Sakaue, K. Yoshikawa, J. Chem. Phys. 117 (2002) 6323.
\bibitem{CPL_Noguchi}H. Noguchi, S. Saito, S. Kidoaki, K. Yoshikawa, Chem. Phys. Lett. 261 (1996) 527.
\bibitem{JCP2_Noguchi}H. Noguchi, K. Yoshikawa, J. Chem. Phys. 109 (1998) 5070.
\bibitem{JCP_Ivanov}V.A. Ivanov, W. Paul, K. Binder, J. Chem. Phys. 109 (1998) 5659.
\bibitem{Macromol_Ivanov}V.A. Ivanov, M.R. Stukan, V.V. Vasilevskaya, W. Paul, K. Binder, Macromol. Theory Simul. 9 (2000) 488.
\bibitem{JCP_Stukan}M.R. Stukan, V.A. Ivanov, A.Yu. Grosberg, W. Paul, K. Binder, J. Chem. Phys. 118 (2003) 3392.
\bibitem{PRE_Stukan}M.R. Stukan, E.A. An, V.A. Ivanov, O.I. Vinogradova, Phys. Rev. E 73 (2006) 051804.
\bibitem{PRE_Schnurr}B. Schnurr, F. Gittes, F.C. MacKintosh, Phys. Rev. E 65 (2002) 061904.
\bibitem{PRE_Miller}I. Miller, M. Keentok, G. Pereira, D. Williams, Phys. Rev. E 71 (2005) 031802.
\bibitem{JCP_Kuznetsov}Y.A. Kuznetsov, E.G. Timoshenko, J. Chem. Phys. 111 (1999) 3744.
\bibitem{JCP_Ou}Z. Ou, M. Muthukumar, J. Chem. Phys. 123 (2005) 074905.
\bibitem{Biophys_Ubbink}J. Ubbink, T. Odijk, Biophys. J. 68 (1995) 54.
\bibitem{Biophys_Odijk}T. Odijk, Biophys. J. 75 (1998) 1223.
\bibitem{Biopolym_Vasilvskaya}V.V. Vasilevskaya, A.R. Khokhlov, S. Kidoaki, K. Yoshikawa, Biopolymers 41 (1997) 51.
\bibitem{JCP_Sakaue}T. Sakaue, J. Chem. Phys. 120 (2004) 6299.
\bibitem{Adv_Yoshikawa}K. Yoshikawa, Y. Yoshikawa, in {\it Pharmaceutical Perspectives of Nucleic Acid-Based Therapeutics}, eds. R.I. Mahato, S.W. Kim, (Taylor $\&$ Francis, London and New York, 2002) 137.
\bibitem{PRL_Murayama}Y. Murayama, Y. Sakamaki, M. Sano, Phys. Rev. Lett. 90 (2003)  018102.
\bibitem{Bio_Damien}D. Baigl, K. Yoshikawa, Biophys. J. 88 (2005) 3486.
\end{thebibliography}
\end{document}